\documentclass[fleqn,10pt]{wlscirep}
\title{The glow of annihilating dark matter \\in Omega Centauri}

\usepackage{textgreek} % enables using greek letters (in body font) outside mathmode, as e.g. \textAlpha
\usepackage{graphicx}

\def\apj{ApJ}
\def\ocen{{\textomega}\,Cen}
\def\rmd{\mathrm{d}}

\usepackage[normalem]{ulem}
\newcommand{\remove}[1]{}

\author[1,2,*]{Anthony M.\ Brown}
\author[1,2,3]{Richard Massey}
\author[4]{Thomas Lacroix}
\author[5]{Louis E.\ Strigari}
\author[3]{Azadeh Fattahi}
\author[6]{C\'eline B{\oe}hm}
\affil[1]{Centre for Extragalactic Astrophysics (CEA), University of Durham, South Road, Durham DH1 3LE, UK}
\affil[2]{Centre for Advanced Instrumentation (CfAI), University of Durham, South Road, Durham DH1 3LE, UK}
\affil[3]{Institute for Computational Cosmology (ICC), University of Durham, South Road, Durham DH1 3LE, UK}
\affil[4]{Instituto de F\'isica  Te\'orica UAM/CSIC, Universidad Aut\'onoma de Madrid, 28049 Madrid, Spain}
\affil[5]{Mitchell Institute for Fundamental Physics and Astronomy, Department of Physics and Astronomy, Texas A\&M University, College Station, TX 77843, USA}
\affil[6]{School of Physics, University of Sydney, Physics Road, Camperdown NSW 2006, Australia}
\affil[*]{anthony.brown@durham.ac.uk}


\begin{abstract}
Dark matter (DM) is the most abundant material in the Universe, but has so far been detected only via its gravitational effects \cite{bert}. Several theories suggest that pairs of DM particles can annihilate into a flash of light at \textgamma-ray wavelengths \cite{louie2018}. While \textgamma-ray emission has been observed from environments where DM is expected to accumulate, such as the centre of our Galaxy \cite{hooper2011,leane2019,gordon2013,ackermann2017}, other high energy sources can create a contaminating astrophysical \textgamma-ray background, thus making DM detection difficult\cite{deboer2017,macias2018}. In principle, dwarf galaxies around the Milky Way are a better place to look -- they contain a greater fraction of DM with no astrophysical \textgamma-ray background -- but they are too distant for \textgamma-rays to have been seen \cite{albert2017}. A range of observational evidence suggests that Omega Centauri ({\textomega}Cen or NGC\,5139), usually classified as the Milky Way's largest globular cluster, is really the core of a captured and stripped dwarf galaxy
\cite{ibata2019,lee1999,sollima2005,piotto2005,bekki2003}. 
Importantly, {\ocen} is ten times closer to us than known dwarfs. Here we show that not only does {\ocen} contain DM with density as high as compact dwarf galaxies, but also that it emits \textgamma-rays with an energy spectrum matching that expected from the annihilation of DM particles with mass 31$\pm$4\,GeV (68\% confidence limit). No astrophysical sources have been found that would otherwise explain {\ocen}'s \textgamma-ray emission, despite deep multi-wavelength searches \cite{henleywillis2018,edwards2001,possenti}. We anticipate our results to be the starting point for even deeper radio observations of {\ocen}. If multi-wavelength searches continue to find no astrophysical explanations, this pristine, nearby clump of DM will become the best place to study DM interactions through forces other than gravity.

\vspace{-3mm}
\end{abstract}
\begin{document}

\flushbottom
\maketitle

\thispagestyle{empty}

\section*{Main}

The thermal DM paradigm suggests a self-annihilation cross section $\langle\sigma v\rangle\sim3$$\times$$10^{-26}$\,cm$^3$\,s$^{-1}$ for particle masses $m_\mathrm{DM}$ in the $10$--$1000$\,GeV range\cite{steigman2012}.
Such particles would annihilate through various channels into \textgamma-ray photons, for which `indirect detection' experiments are searching. 
Dwarf spheroidal (dSph) galaxies are the ideal places to look, because they are made of mainly DM, and contain few astrophysical sources that also emit at \textgamma-ray energies. A study of 43 dSphs, using 6\,years of observations from the \textit{Fermi} satellite's Large Area Telescope (\textit{Fermi}-LAT), found no evidence of \textgamma-ray emission %from DM annihilation 
\cite{albert2017}. 
However, the observed dSphs were all more than 23\,kpc away (100\,kpc on average), and the \textgamma-ray flux reaching Earth falls as the inverse square of this distance. Since the cross section is not known, the flux may simply be too weak to have been detected.

Indirect detection experiments have \textit{not} looked at globular clusters, because they contain little DM; they are %gravitationally bound, 
spherical collections of stars that formed from a single gas cloud early in the life of a galaxy. But of all the Milky Way's $\sim200$ globular clusters, Omega Centauri ({\ocen}; RA $201.7^\circ$, Dec $-47.56^\circ$) is unique. It is the most massive\cite{baumgardt2017}, the most luminous\cite{harris1996}, and has the largest core and half-light radius\cite{harris1996}. {\ocen} also possesses multiple stellar populations with a large spread in metallicity and  spatial distributions\cite{lee1999,sollima2005,piotto2005} that include a trailing stellar stream \cite{ibata2019}. Even its orbit around the Milky Way is retrograde\cite{dinescu1999}. This growing list of atypical properties suggests that {\ocen} may not be a globular cluster, but the remnant core of a dSph, whose outskirts were tidally stripped as it fell into the Milky Way \cite{bekki2003}. Since {\ocen} is currently only 5.4\,kpc from Earth\cite{Weldrake2007}, if it were indeed a dSph, it would be the best place to look for annihilating DM.

\begin{figure}[p]
\centering
\includegraphics[width=100mm]{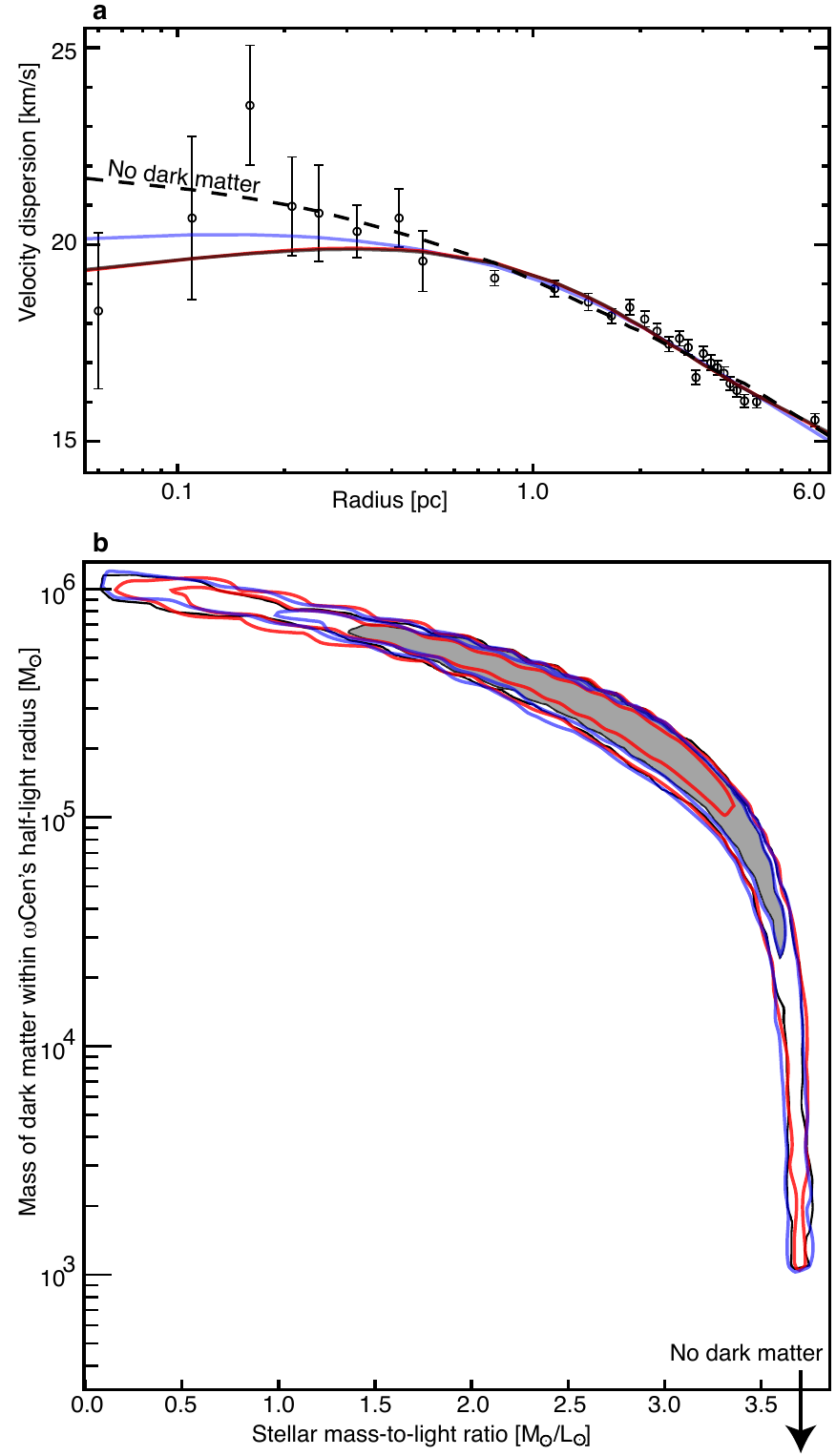}
\caption{{\ocen} contains a spatially extended component of DM. 
\textbf{Panel a} shows the tangential velocity dispersion of stars moving within {\ocen}'s gravitational potential, as a function of radius from its centre, reproduced from Watkins et al.\ (2015)~\cite{Watkins2015}. Models that include DM (red: Moore, black: NFW, blue: Burkert) are a better fit than a model without DM (dashed).
\textbf{Panel b} shows $68\%$ and $95\%$ confidence limits on the mass of dark matter within the central 7\,pc, and the stellar mass-to-light ratio, the parameter with which it is most correlated. Models without DM (off the bottom of this plot) are excluded with $p<0.05$.
}
\label{fig:MLDM}
\end{figure}

We have measured the distribution of mass inside {\ocen}, and find that it does indeed contain a component of DM.
We infer this result from the observed velocities of its stars (the tangential component of which is shown in figure~\ref{fig:MLDM}a), under the assumption that they are in dynamic equilibrium with the gravitational potential. 
We model the gravitational potential as three components of mass: luminous stars, a central black hole, and non-luminous DM.
Previous analyses have disagreed over whether the data require a black hole \cite{baumgardt2017} or not \cite{vdM2010}.
We find at most a low mass $M_\mathrm{BH}<5\times10^{4}\,\mathrm{M_\odot}$ black hole. However, the data are always better fit ($p<0.05$ for no DM) if we include an extended distribution of DM (figure~\ref{fig:MLDM}b).
The best-fit model using a Navarro-Frenk-White distribution of DM contains $10^{5.3\pm0.5}$\,M$_\odot$ of DM inside {\ocen}'s optical half-light radius, 7\,pc.
This is about half of the corresponding stellar mass, and robust to changes in the assumed DM density profile, or omission of the central black hole (see Methods).
 
\begin{figure}[p]
\centering
\includegraphics[width=89mm]{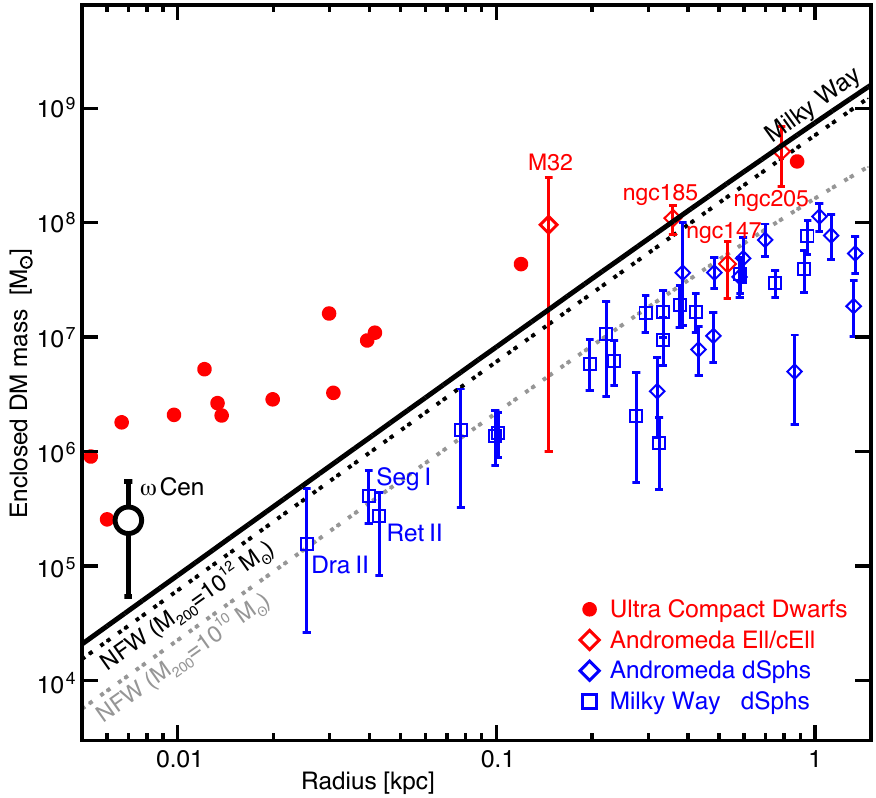}
\caption {
{\ocen} is an ideal target for DM indirect detection experiments, being nearby and containing a higher density of DM than most dSph galaxies (and even the centre of the Milky Way) although it is truncated and contains less DM in total. It most resembles compact dwarf galaxy M32, likely the remnant core of a massive galaxy that was tidally stripped by the Milky Way's neighbouring Andromeda galaxy with contracted DM distribution. Points show the mass of DM contained by dwarf galaxies of the Local Group and ultra-compact dwarfs in Fornax, within radii at which measurements are available.}

\label{fig:DMmass}
\end{figure}

The best-fit DM distribution in {\ocen} is remarkably compact. Its geometric $J$-factor, which determines how many DM particles get close enough to potentially annihilate, is $\log_{10}(J\,[\mathrm{GeV}^{2}\,\mathrm{cm}^{-5}])=22.1^{+1.3}_{-0.9}$ ($68$\% confidence limit).  
This is larger than most dSphs, e.g.\ Segue~I, the nearest in the \textit{Fermi}-LAT sample at 23\,kpc, has $J=10^{19.4\pm0.3}\,\mathrm{GeV}^{2}\,\mathrm{cm}^{-5}$. 
Kinematically (figure~\ref{fig:DMmass}) and on the size-luminosity relation\cite{Tolstoy2009}, {\ocen} most resembles Ultra Compact Dwarf galaxies (UCDs) and the Local Group compact dwarf galaxy M32\footnote{M32 is the only known compact dwarf galaxy in the Local Group, and one of Andromeda's brightest satellites. The realisation that the Milky Way may possess a similarly compact satellite %like M32 
would resolve much of the unexplained asymmetry in the satellite populations of these sister galaxies.}, which is thought to be the core remnant of a massive galaxy that was tidally stripped as it fell into  the Milky Way's neighbour Andromeda\cite{dsouza2018}.  
The high density of DM in compact dwarfs can be generated by contraction in the gravitational potential of a dominant compact baryonic component\cite{Abadi2010}.

\begin{figure}[p]
\centering
\includegraphics[width=100mm]{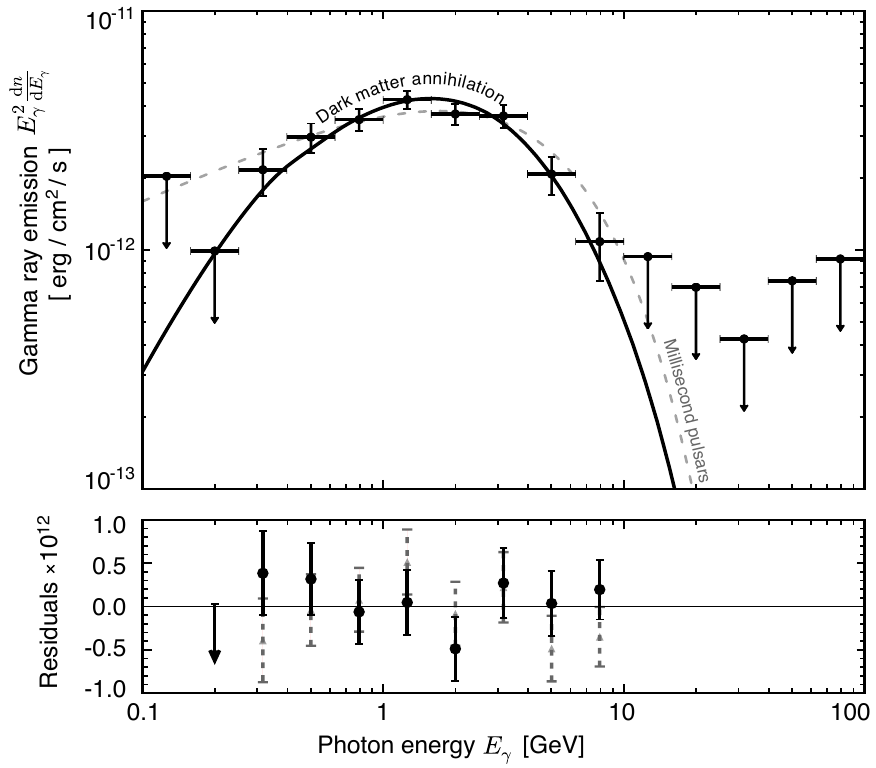}
\caption{{\ocen} emits \textgamma-rays, and their energy spectrum is consistent with DM annihilation via the $b\bar{b}$ channel (solid line). 
It is a poor match to the spectral shape of local MSPs (dashed line).
Data points show the energy flux detected by \textit{Fermi}-LAT, integrated over 10\,years. Error bars show $\pm1\sigma$ statistical uncertainties, or $2\sigma$ upper limits for those spectral bins with detection test statistic TS $<20$.
}
\label{fig:spec}
\end{figure}

Given all the evidence that nearby {\ocen} contains DM, it might be possible to detect \textgamma-ray emission from its annihilation -- if the DM particle mass happens to be in an appropriate mass range, and the annihilation cross section is sufficiently high. 
To test this hypothesis, we have integrated all observations from the first ten years of \textit{Fermi}-LAT operations. After accounting for diffuse \textgamma-ray emission and known nearby sources, we find that {\ocen} is indeed a bright \textgamma-ray source, from which we detect an energy flux $1.13\pm0.05 \times 10^{-11}$\,ergs\,cm$^{-2}$\,s$^{-1}$ that does not vary over time. 
Assuming that the emission is isotropic, its total luminosity is $3.94 \pm 0.17 \times 10^{34}$ ergs s$^{-1}$ in the $0.1$--$100$\,GeV energy range, with the spectral energy distribution shown in figure~\ref{fig:spec}.

Could prosaic astrophysical sources be responsible for {\ocen}'s \textgamma-ray emission?
The \textgamma-ray emission from globular clusters is usually attributed to millisecond pulsars (MSPs): neutron stars spun up by close encounters with a giant star that overflowed its Roche lobe and accreted material that was subsequently radiated as \textgamma-rays, X-rays and radio waves. However, no MSPs are \textit{expected} inside {\ocen}. If {\ocen} is a globular cluster, its stellar encounter rate is sufficiently low that none are expected to form \cite{bahramian2013}.
If {\ocen} is a dSph, its stellar encounter rate is orders of magnitude lower still \cite{santana2013}.
Moreover, no MSPs are \textit{observed} inside {\ocen}. 
While possible MSP candidates have been identified with deep X-ray observations based purely on their position in the X-ray colour-luminosity diagram\cite{henleywillis2018}, the Parkes Radio Telescope survey of globular clusters targeted its deepest observations at {\ocen}\cite{edwards2001} and found none\cite{possenti} (see Methods for more information). 
Alternative possible sources of \textgamma-rays include cataclysmic variable stars, but only those within our immediate galactic neighbourhood have been observed to be \textgamma-ray bright\cite{cv}; or magnetospheric \textgamma-ray emission from the vicinity of a black hole, but that is variable %\cite{katsoulakos2018}, 
and would be too faint to detect, given the low mass of any black hole in {\ocen}\cite{nerenov2008}.

If DM annihilation is responsible for {\ocen}'s \textgamma-ray emission, we can infer DM particle properties from the \textgamma-ray spectral energy distribution. The observed spectrum is well-fit by a model of DM annihilation into $b$ and $\bar{b}$ quarks, followed by prompt photon emission. % \cite{Cirelli2011}. 
We find a best-fit DM particle mass (which determines the energy scale of peak emission) $m_\mathrm{DM}=31\pm 4$\,GeV, and velocity-averaged annihilation cross section (which determines the flux amplitude) $\mathrm{log}_{10}(\langle\sigma v\rangle \,[\mathrm{cm}^3\,\mathrm{s}^{-1}])\,=-28.2^{+0.6}_{-1.2}$ (68\% confidence limits, assuming a Navarro-Frenk-White distribution of DM plus a black hole). 
Our best-fit DM mass is consistent with other indirect detection experiments that all, individually, provide tentative evidence for DM annihilation \cite{hooper2011,leane2019,gordon2013}. 
Importantly, our best-fit annihilation cross section is consistent with the $-26.4$ upper limit derived from the non-detection of \textgamma-rays recently reported by the \textit{Fermi}-LAT collaboration from a stacked analysis of more distant dSphs\cite{albert2017}. 

Diverse observational evidence thus points to {\ocen} being the stripped core of a dwarf galaxy, and supports an interpretation of its observed \textgamma-ray spectrum as originating from annihilating DM. That our best-fit DM particle parameters are consistent with other tentative results suggests that there is nothing unique about the DM within {\ocen}, and that the DM annihilation signal is detectable in {\ocen} simply because of its proximity to Earth. 
Its DM is more dense than that in the Galactic centre.
Its $J$ factor is not as high ($J\sim10^{23.1}\,\mathrm{GeV}^{2}\,\mathrm{cm}^{-5}$)\cite{Boddy2018} because it is not as large, but it has no known sources of background \textgamma-ray emission. 
If multi-wavelength searches continue to find no astrophysical explanations for the \textgamma-ray emission, this pristine, nearby clump of DM will become the best place to study the nature of DM.

\section*{Acknowledgements}
We acknowledge the excellent data and analysis software provided by the \textit{Fermi}-LAT collaboration. A.M.B.\ , R.M.\ and A.F. acknowledge financial support from the UK Science and Technology Facilities Council, grant ST/P000541/1. 
T.L.\ has received financial support from CNRS-IN2P3 and the University of Montpellier, and from the European Union’s Horizon 2020 research and innovation programme under Marie Sk\l{}odowska-Curie grant agreement No.\ 713366. The work of T.L.\ has also been supported by the Spanish Agencia Estatal de Investigaci\'{o}n through grants PGC2018-095161-B-I00, IFT Centro de Excelencia Severo Ochoa SEV-2016-0597, and Red Consolider MultiDark FPA2017-90566-REDC.
R.M.\ is supported by a Royal Society University Research Fellowship, and A.F.\ by an EU COFUND/Durham Junior Research Fellowship under EU grant agreement no.\ 609412. 

\section*{Author contributions statement}
A.M.B.\ and R.M.\ conceived the experiment. L.E.S.\ and A.F.\ analysed stellar kinematics to fit the amount of DM. A.F.\ performed the analysis and comparison with dwarf galaxies. A.M.B.\ analysed the \textit{Fermi}-LAT data to measure the \textgamma-ray emission. T.L.\ and C.B.\ interpreted the DM particle physics implications of the \textgamma-ray signal. All authors reviewed the manuscript. 

\section*{Additional information}
The authors declare no competing interests.
%To include, in this order: \textbf{Accession codes} (where applicable); \textbf{Competing financial interests} (mandatory statement). 

%The corresponding author is responsible for submitting a \href{http://www.nature.com/srep/policies/index.html#competing}{competing financial interests statement} on behalf of all authors of the paper. This statement must be included in the submitted article file.

%%%%%%%%%%%%%%%%%%%%%%%%%%%%%%%%%%%%%%%%%%%%%%%%%%%%%%%%%%%%%%%%%%%%%%%%
~
\newpage
\section*{Methods}
\setcounter{section}{0}
%%%%%%%%%%%%%%%%%%%%%%%%%%

%%%%%%%%%%%%%%%%%%%%%%%%%%%%%%%%%%%%%%%%%%%%%%%%%%%%%%%%%%%%%%%%%%%%%%%%%%%%%%%%%%%%%%
\section{Distribution of dark matter, measured via stellar kinematics} \label{density_profile}

To infer the distribution of matter, we exploit the spherical Jeans equation
\begin{equation}
\frac{\rmd \left(\rho_\star \, \sigma_{\star,r}^2 \right)}{\rmd r} + \frac{2 \beta}{r} \rho_\star \, \sigma_{\star,r}^2 + \rho_\star \frac{\rmd\Phi}{\rmd r} = 0,
\label{eq:jeans2}
\end{equation}
where $r$ is the 3D radius, $\sigma_{\star,r}^2(r)$ is the radial velocity dispersion of the stars, $\beta= 1 - \sigma_t^2/\sigma_r^2$ is the anisotropy of their orbits, $\rho_\star(r)$ is the stellar density profile, and $\Phi(r)$ is the total gravitational potential of all matter. The 3D velocity dispersion of the stars can be split into one observable component along the line-of-sight and two (radial and tangential) components of `proper motion' in the plane of the sky
\begin{eqnarray} 
\sigma_{\star, {\rm LOS}}^2(R) &=& \frac{2}{L_\star(R)} \int_R^\infty \left[1-\beta\frac{R^2}{r^2}\right] \frac{r \rho_\star \sigma_{\star,r}^2}{\sqrt{r^2-R^2}} ~\rmd r \\
\sigma_{\star, \rm R}^2(R) &=& \frac{2}{L_\star(R)} \int_R^\infty \left[1-\beta+\beta\frac{R^2}{r^2}\right] \frac{r \rho_\star \sigma_{\star,r}^2}{\sqrt{r^2-R^2}} ~\rmd r \\
\sigma_{\star, \rm T}^2(R) &=& \frac{2}{L_\star(R)} \int_R^\infty \left[1-\beta\right] \frac{r \rho_\star \sigma_{\star,r}^2}{\sqrt{r^2-R^2}} ~\rmd r~, \label{eq:sigmat} 
\label{eq:projectedjeans} 
\end{eqnarray} 
where $R$ is the 2D projected radius.

Measurements of the line-of-sight velocity dispersion in {\ocen} have been obtained using VLT/MUSE \cite{Kamann2018} and Keck \cite{Baumgardt2018}. Proper motions have been measured with \textit{Gaia} \cite{Baumgardt2019} and the \textit{Hubble Space Telescope} (\textit{HST}) \cite{Watkins2015}. 
\textit{Gaia} cannot measure crowded fields inside $R=4.7$\,pc, but we find that the DM density at low $r$, and hence the $J$-factor, is best constrained using measurements that cover scales smaller than this. As our primary data\footnote{Using instead proper motions on larger scales from \textit{Gaia} DR2, or line-of-sight velocities from MUSE, gives statistically consistent but far noisier results. We avoid combining line-of-sight velocity measurements with proper motions for constraining variations in $\beta(r)$ as a function of radius. The reason is that the two measurements correspond to different stellar populations which can have different kinematical properties, as seen e.g.\ in Fornax and Sculptor dwarf galaxies\cite{Walker2011}.
}, we therefore use tangential velocity dispersions from the high resolution \textit{HSTpromo} survey\cite{Watkins2015} (figure~\ref{fig:MLDM}a). 
These data have not previously been analysed in the context of DM.
Because measurements are confined to stars more luminous than $\sim1$ magnitude below the main sequence turnoff, we do not need to account for the effect of mass segregation or changes in the stellar mass-to-light ratio.  
Furthermore, because \textit{HST} resolves individual stars, adjacent data points are uncorrelated (unlike line-of-sight velocities derived from the width of emission lines, which are subject to beam smearing). When calculating goodness of fit statistics, we assume that log-likelihoods on $\sigma_{\star, \rm T}^2(R)$ are uncorrelated, symmetric Gaussians with $1\sigma$ dispersions given by the uncertainties in Watkins et al.\ (2015). 
To convert observed quantities (projected angular distances and proper motions) to physical quantities, we assume {\ocen} to be $d=5.4\,\mathrm{kpc}$ from the Earth, from its distance modulus corrected for reddening \cite{Weldrake2007}.

Assumptions in equation~\eqref{eq:jeans2} include spherical symmetry. This is reasonable even if {\ocen} has been affected by tides, because tidal stripping makes DM distributions more spherical\cite{barber2016}. Star formation in the centres of DM halos also decreases triaxiality\cite{Chua2019}. Equation~\eqref{eq:jeans2} assumes dynamical equilibrium. This is reasonable too: objects orbiting the Milky Way are tidally shocked and out of equilibrium close to the pericentre of their orbit, but they reach equilibrium shortly after due to the relatively short time-scale of their crossing times\cite{Penarrubia2009} ($\sim$$1$\,Myr for {\ocen}). We note that, according to the latest proper motions measurements from the Gaia mission, \ocen\ is currently close to the apocentre of its orbit\cite{Helmi2018,Baumgardt2019}.

We model {\ocen}'s gravitational potential $\Phi(r)$ as the sum of three components: luminous stars, a central black hole, and non-luminous DM. We infer the 3D distribution of stellar mass $\rho_\star(r)$ by de-projecting the surface brightness distribution at visible wavelengths, $L_\star(R)$, assuming spherical symmetry and a mass-to-light ratio $M_\star/L_\star$ that is constant but free to vary. We use measurements of $L_\star(R)$ from a fit\cite{vdM2010} that included different stellar types, so for our purposes it may be thought of as an average light profile for the stars. 
We add a black hole as a central point mass.
We add DM in a Navarro-Frenk-White\cite{nfw1996} (NFW) profile
\begin{equation}
\label{NFW}
\rho_{\mathrm{NFW}}(r) = \rho_{\rm s} \left( \dfrac{r}{r_{\rm s}} \right)^{-1} \left(1 + \dfrac{r}{r_{\rm s}} \right)^{-2},
\end{equation}
or, to explore different effects of adiabatic contraction and gasdynamic feedback, either a cuspier Moore\cite{moore1999} profile
\begin{equation}
\label{Moore}
\rho_{\mathrm{Moore}}(r) = \rho_{\rm s} \left( \dfrac{r}{r_{\rm s}} \right)^{-1.5} \left(1 + \dfrac{r}{r_{\rm s}} \right)^{-1.5},
\end{equation}
or a cored Burkert\cite{burkert1995} profile 
\begin{equation}
\label{Burkert}
\rho_{\mathrm{Burkert}}(r) = \rho_{\rm s} \left( 1+\dfrac{r}{r_{\rm s}} \right)^{-1} \left(1 + \dfrac{r^2}{r_{\rm s}^2} \right)^{-1},
\end{equation}
where $\rho_{\rm s}$ and $r_{\rm s}$ are the scale density and radius, respectively. 
In all three cases, we assume the DM is truncated at a tidal radius\cite{vdM2010} $r_{\rm t}=65$\,pc.

We use a nested sampling algorithm\cite{multinest} to fit the model's free parameters, adopting flat priors on the stellar mass-to-light ratio $M_\star/L_\star$ and velocity anisotropy $\beta$. For models with dark matter, we also fit free parameters for its scale radius $r_{\rm s}$ and total mass $M_\mathrm{DM}$ (which is proportional to $\rho_{\rm s}$), adopting flats priors in log space. To fit models with a black hole, we fit its mass $M_\mathrm{BH}$, adopting a flat prior in log space. We find that the stellar kinematic data are broadly insensitive to the slope of the DM's inner density profile ($\chi^2_\mathrm{Burkert}=36$, $\chi^2_\mathrm{NFW}=38$, $\chi^2_\mathrm{Moore}=39$ in 23 degrees of freedom). Adding a black hole changes these by $<1\%$, and its only effect is to compensate for the difference in cuspiness between an NFW and Moore DM profile, by moving the maximum likelihood position along the degeneracy between $M_\mathrm{DM}$ and $M_\star/L_\star$. We therefore adopt a baseline model with a black hole and NFW distribution of dark matter. Median and (1D marginalised) 68\% confidence limits for this model's parameters are 
$M_\star/L_\star=2.75^{+0.68}_{-0.96}\,M_\odot/L_\odot$, 
$\beta=0.08^{+0.18}_{-0.29}$ for mildly radial orbits, 
$\mathrm{log}_{10}(M_\mathrm{BH}\,[M_\odot])=3.93^{+0.73}_{-0.63}$,
$r_{\rm s}=1.63^{+3.00}_{-1.52}$\,pc, and integrated total DM mass
$\mathrm{log}_{10}(M_\mathrm{DM}\,[M_\odot])=5.40^{+0.34}_{-0.67}$.
Removing all DM gives a best-fit model with a physically disfavoured\cite{Conroy2009} $M_\star/L_\star=3.73^{+0.02}_{-0.02}\,M_\odot/L_\odot$, and statistically disfavoured $\chi^2=49$. 
Integrating under the marginalised log-likelihood distribution, we rule out the hypothesis that {\ocen} contains no DM, with $p<0.05$. 

\begin{figure}[tb]
\centering
\includegraphics[width=89mm]{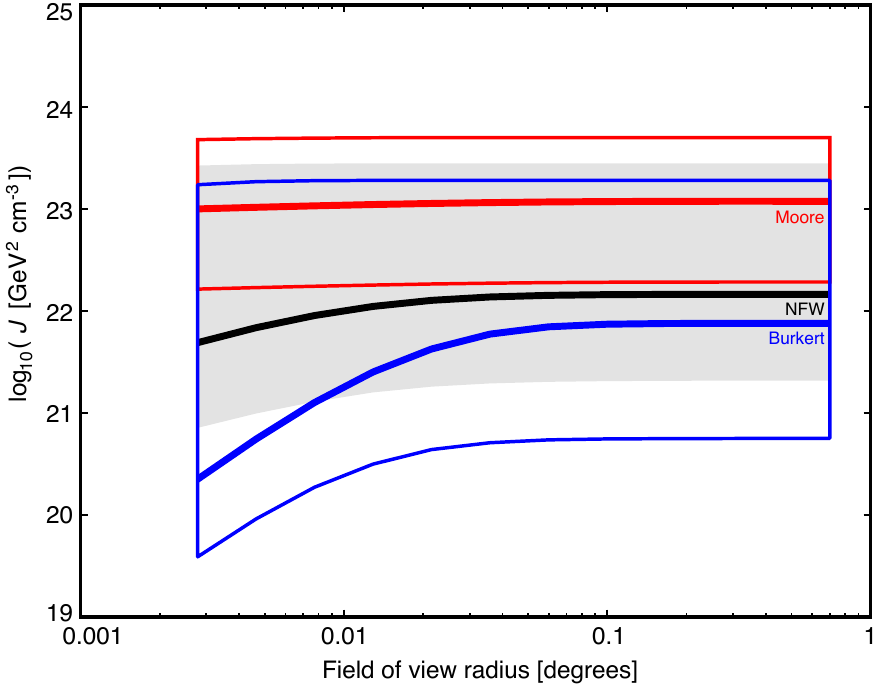}
\caption{Constraints on the $J$-factor of {\ocen}, the squared density of its DM along a line of sight. 
Along with the interaction cross section, this determines the total number of photons produced by DM annihilation. 
Thick lines show the median, and boxes show 68\% containment regions of the posterior probability distribution for a model with DM in a (red: Moore, black: NFW, blue: Burkert) profile.}
\label{fig:jfactor}
\end{figure}

Because self-annihilation requires two particles, the DM annihilation rate depends on its density squared. The integral of this along a line of sight and over a solid angle $\Delta\Omega$, is known as the $J$-factor
\begin{equation}
J (\Delta\Omega) = \dfrac{1}{4\pi} \int_{\Delta \Omega} \int_{\mathrm{l.o.s.}} \! \rho^{2}_{\mathrm{NFW}}\left(r\left(s,\Omega\right)\right)  \, \mathrm{d}s \, \mathrm{d}\Omega,
\end{equation} 
where $s$ is the line of sight coordinate, $\mathrm{d}\Omega = 2 \pi \sin \Psi \, \mathrm{d} \Psi$ in spherical symmetry with $\Psi$ the angular radius from the center of {\ocen}, and
\begin{equation}
r(s,\Omega) = \sqrt{s^2 + d^2 - 2 s d \cos \Psi}~,
\end{equation} 
where $d = 5.4\,\mathrm{kpc}$, $r_{\rm t}=65$\,pc.
We perform the integration over an aperture of radius $\Delta\Omega=0.7^\circ$, and along the line of sight from $s_{\rm min} = d \cos \Psi - \sqrt{r_{\rm t}^2 -d^2 \sin^2 \Psi}$ to $s_{\rm max} = d \cos \Psi + \sqrt{r_{\rm t}^2 -d^2 \sin^2 \Psi}$. All the density profiles we consider fall off sufficiently quickly at large radii that $J$ does not increase if either $\Delta\Omega$ or $r_{\rm t}$ are increased (figure~\ref{fig:jfactor}). 

We propagate the entire posterior probability distribution of $\rho_{\mathrm{DM}}(r)$, via a Markov Chain, to a posterior probability distribution on $\log_{10}(J\,[\mathrm{GeV}^{2}\,\mathrm{cm}^{-5}])=22.1^{+1.3}_{-0.9}$, where we quote the median values and $68\%$ confidence limits. If we instead assume a Burkert or Moore profile, this value is $21.9^{+1.4}_{-1.1}$ or $23.1^{+ 0.6}_{- 0.8}$.

%%%%%%%%%%%%%%%%%%%
\section{Dark matter content of other dwarf galaxies}

To compare the enclosed DM mass of dwarf galaxies with that of {\ocen} (figure~\ref{fig:DMmass}), we do not perform Jeans modelling of their mass distribution, because we do not need to constrain their full radial profile. Moreover more distant dwarfs (e.g. UCDs) lack detailed kinematic measurements for Jeans modelling. 
The total mass within the 3D half-light radius ($r_{\rm h}$) of dispersion-supported systems (dSphs and ellipticals) is well constrained using their stellar velocity dispersion. We use the following formula\cite{Wolf2010}
\begin{equation}
M_{\rm tot}(<r_{\rm h})=3G^{-1}\sigma_{\rm los}^2r_{\rm h}~,
\end{equation}
where $\sigma_{\rm los}$ is the luminosity weighted velocity dispersion of stars. We compile the data for all known dwarf galaxies within $300$kpc from either Milky~Way or Andromeda in the Local Group\cite{McConnachie2012,Fattahi2018}, as well as UCDs in Fornax \cite{Mieske2008}. 
After obtaining $M_{\rm tot}(<r_{\rm h})$ for dwarfs with available $\sigma_{\rm los}$ and $r_{\rm h}$ measurements, their enclosed DM mass is derived by subtracting the stellar mass from the total mass within $r_{\rm h}$. Stellar masses are calculated from luminosities\cite{McConnachie2012} and stellar mass-to-light ratios\cite{Woo2008} in the V-band. 

The mean DM density within the half-light radius of {\ocen}, 7pc, is higher than at the centre of the Milky Way\cite{Sofue2013}, or an NFW profile with $M_{200}=10^{12}\,M_{\odot}$ (values of the concentration parameter for the NFW profiles in figure~\ref{fig:DMmass} are based on the average $M_{200}$-concentration relation in $\Lambda$CDM cosmological simulations\cite{Ludlow2016} with Planck cosmological parameters). 
The DM in {\ocen} is also more dense than in dSphs of various stellar mass (e.g.\ Segue~I, Draco, Fornax with $M_{\star}\sim10^2, 10^5, 10^7 \, M_{\odot}$, respectively) which are all consistent with living in DM halos with $M_{200}\sim 10^{10}\,M_{\odot}$ or lower.
{\ocen} is most similar to the compact dwarf elliptical satellite of Andromeda galaxy, M32. 
Constraining the central DM mass in M32 is difficult due to the internal kinematics being dominated by the stellar component and central black hole\cite{Howley2013} (like in {\ocen} but more distant); kinematics of stars in the outer parts of M32, however, point to an extended DM halo.
Indeed, the DM density in {\ocen} is approaching that of (ultra) compact dwarf galaxies\cite{Goerdt2008} (shown on the plot assuming $M/L=1.6\,M_\odot/L_\odot$), unless their stellar mass-to-light ratio is as high as $\sim5$. 
Previous speculation that such a value is unphysically high (twice that in globular clusters), led to debates about uncertainties in stellar population models or stellar initial mass function\cite{Mieske2008}, rather than the possible presence of DM.

\begin{table}[tb]
   \caption{Selection criteria used to extract observations of {\ocen} from the \textit{Fermi}-LAT database.
   \vspace{-2mm}
   \label{tab:selectioncuts}}
   \begin{center}
   \begin{tabular}{lc} 
      \hline \hline 
      Science Tools version     & \textsc{11-04-00}  \\ 
      Instrument Response Function & \textsc{p8r3\_source\_v2}    \\ 
      Event class               & \textsc{source (front$+$back)}, Pass 8   \\
      							& \textsc{evclass=128} \\
      Photon Energies           & $0.1$--$100$\,GeV    \\ 
      MET$_\mathrm{start}$ 				& 239557417		\\ 
      MET$_\mathrm{stop}$				& 555090221  \\
      RoI                       & $15^{\circ}$      \\
      %Model RoI 				& $25^{\circ}$      \\
      Zenith angle cut          & $<90^{\circ}$    \\  
      Data quality              & $>0$ \\
      LAT config                & $1$   \\
      %Galactic diffuse model    & \texttt{gll\_iem\_v07.fit} \\
      %Isotropic diffuse model   & \texttt{iso\_P8R3\_SOURCE\_V2\_v1.txt} \\ 
      \hline \hline 
    \end{tabular}
    \end{center}
  \label{analysis}
\end{table}

\begin{table*}[t]
\caption{Newly identified point sources emitting \textgamma-rays with detection with $\sigma > 5$ significance, within $15^\circ$ of {\ocen}. These were not present in the 4FGL catalogue, which was derived from the first eight years of \textit{Fermi}-LAT observations. Our study combines the first ten years of \textit{Fermi}-LAT observations, with an extra 25\% of live-time. }
\centering
\begin{tabular}{lllll}
\hline\hline
Point Source & RA (J2000) & Dec. (J2000) & Offset & TS \\ 
 & (degrees) & (degrees) & (degrees) & \\
\hline
PS J1321.1-4527 & 200.278 & -45.454 & 2.235 & 105.0 \\
PS J1357.5-4907 & 209.394 & -49.118 & 5.380 & 31.1 \\
PS J1338.1-4200 & 204.536 & -42.001 & 5.823 & 30.9 \\
PS J1402.0-4645 & 210.513 & -46.755 & 6.040 & 54.3 \\
PS J1401.3-5012 & 210.332 & -50.208 & 6.307 & 45.3 \\
PS J1430.7-5537 & 217.678 & -55.618 & 12.804 & 32.8 \\
\hline\hline
\end{tabular}
\label{tab:NewSources}
\end{table*}

\begin{figure}[t]
\centering
\includegraphics[width=89mm]{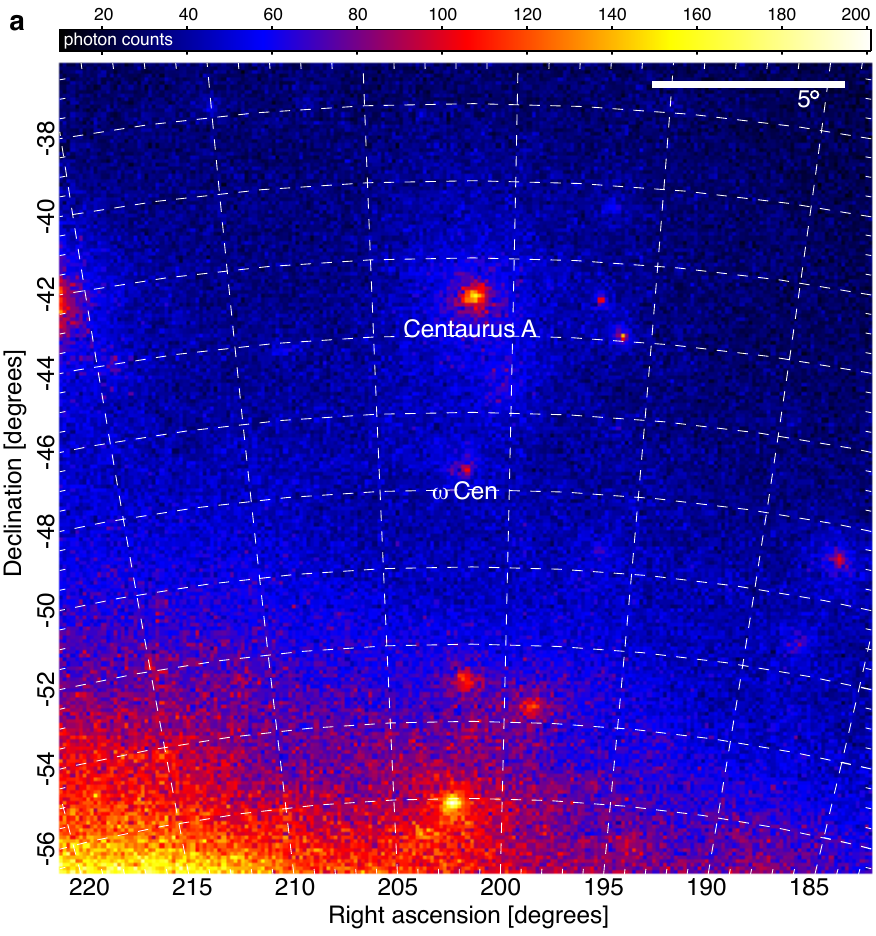}~
\includegraphics[width=89mm]{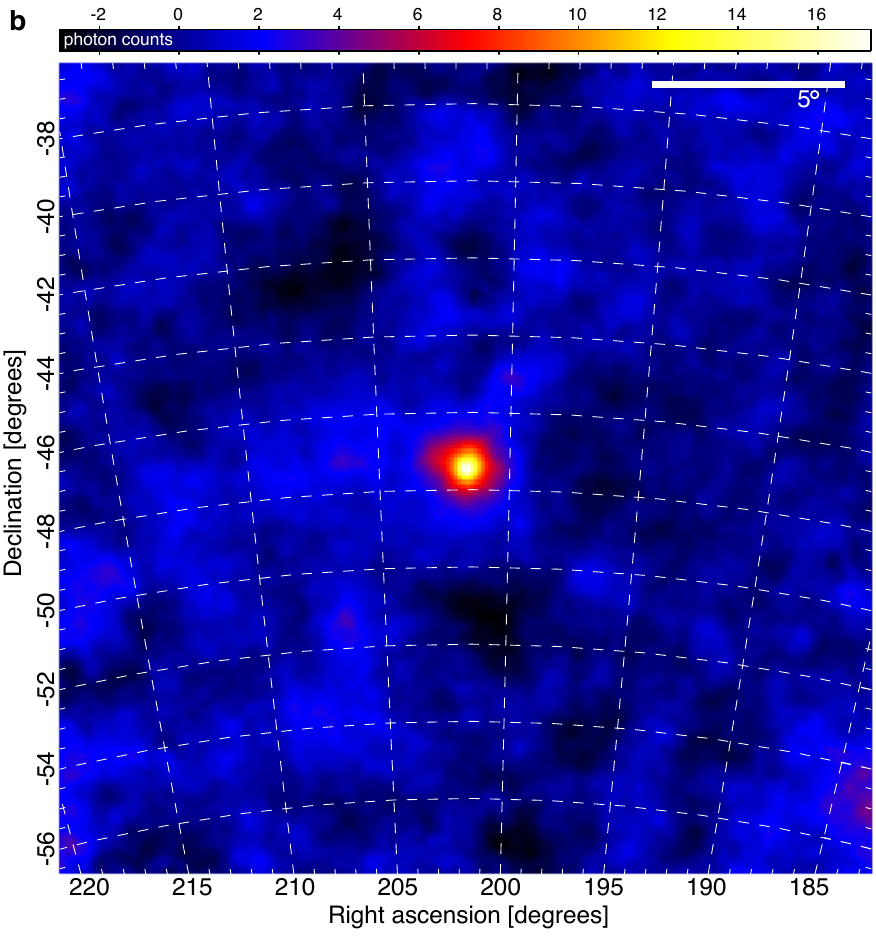}
\caption{Map of \textgamma-ray emission near {\ocen}. Colour scales represent the number of $0.1$--$100$\,GeV photons per $0.1^{\circ}\times0.1^{\circ}$ pixel, during the 10-year {\em Fermi} mission.
\textbf{Panel a} shows all detected photons. The bright region at the bottom left is diffuse \textgamma-ray emission from the Galactic plane. \textbf{Panel b} shows the (statistically insignificant) residuals once all sources of \textgamma-rays except {\ocen} have been modelled and subtracted. The bright excess at the centre are the statistically significant photons associated with {\ocen} .
}
\label{fig:map}
\end{figure}

\begin{figure}[tb]
\centering
\includegraphics[width=1.0\linewidth]{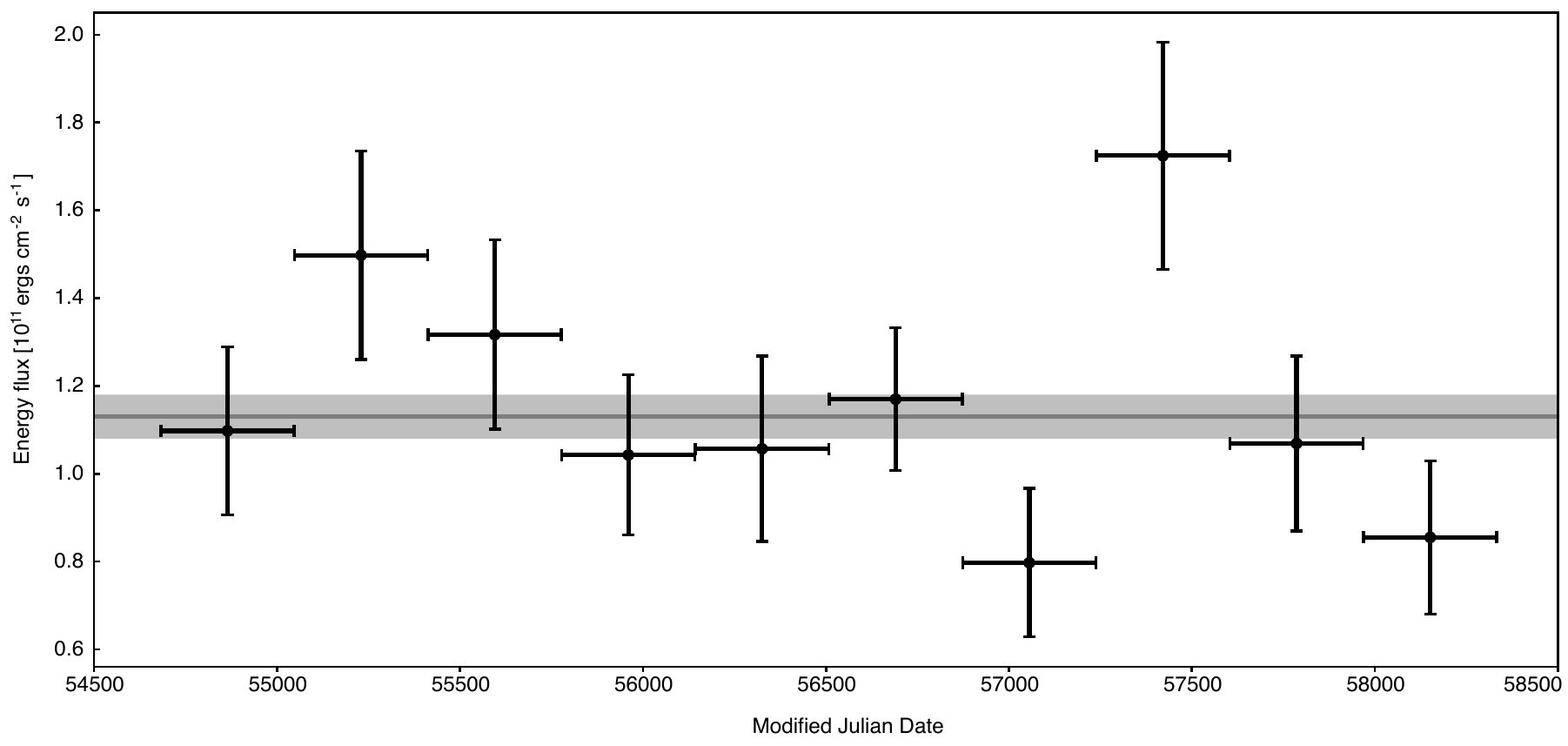}
\caption{
The \textgamma-ray flux from {\ocen} has shown no (statistically significant) sign of variability over the 10\,years of the \textit{Fermi} mission.
Data points show the mean and $1\sigma$ statistical uncertainty of energy flux in $1$--$1000$\,GeV photons, binned in $1$\,year intervals. Flux from {\ocen} is significantly ($\sigma>7$) detected in each individual bin. A likelihood analysis finds this time series to be consistent with a constant flux: the mean and $1\sigma$ statistical uncertainty of which is shown as a grey band.
\label{fig:lc}
}
\end{figure}

%%%%%%%%%%%%%%%%%%%%%%%%%%%%%%%%%%%%%%%%%%%%%%%%%%%%%%%%%%%%%%%%%%%%%%%%%%%%%%%%%%%%%%
\section{Gamma-ray emission observed by \textit{Fermi}-LAT}

We analyse all spacecraft and event data collected by \textit{Fermi}-LAT during the first ten years of its science mission, from 2008 August 4 to 2018 August 4 (Mission Elapsed Time (MET) period of 239557417\,s to 555090221\,s). Its point-spread function (95\% containment) is a maximum of $\sim12^\circ$ for the lowest photon energy $E_{\gamma}=0.1$ GeV. We therefore collate all $0.1<E_{\gamma}<100$ GeV \textsc{source (front$+$back)} events within a $15^{\circ}$ radius of interest (RoI) centred on {\ocen}. Following Pass~8 data analysis guidelines and our previous analysis of other globular clusters\cite{brown2018}, we apply a $90^{\circ}$ cut on the zenith angle of each event to remove \textgamma-rays that originated from cosmic ray-induced air showers in the Earth's atmosphere. Then, we select good time intervals by applying a `\textsc{data\_qual$>$0 \&\& lat\_config==1}' filter criterion to the resulting data set (figure~\ref{fig:map}a). The event selection criteria used in our analysis are summarised in table~\ref{tab:selectioncuts}. 

We model the \textgamma-ray flux as a sum of discrete, point-like sources (including {\ocen} itself), two extended sources, and two components of diffuse emission: 
\begin{itemize} \itemsep0em
    \item 
All 368 point-like sources in the \textit{Fermi}-LAT 8-Year Point Source Catalog (4FGL)\cite{4fglcat}, within $25^{\circ}$ of {\ocen}.     \item 
The largest extended structure is lobe emission from the jets of radio galaxy Centaurus A (4FGL J1324.0-4330e). We define their spatial extent using the \texttt{CenALobes.fits} template provided by the \textit{Fermi}-LAT collaboration, and a power law spectrum with the spectral index frozen to the 4FGL value. The other extended object is FGES J1409.1-6121 (4FGL J1409.1-6121e), which we model as a disc of radius $0.733^{\circ}$ and log-parabolic spectrum. 
    \item 
Isotropic diffuse \textgamma-ray emission is defined by \textit{Fermi's}  \texttt{iso\_\textsc{P8R3}\_\textsc{SOURCE}\_\textsc{V2}\_\textsc{V1}.txt} tabulated spectrum, with normalisation free to vary. Galactic diffuse emission is modeled with \textit{Fermi}'s \texttt{gll\_iem\_v07.fits} spatial map and a power-law spectrum with normalisation free to vary.
\end{itemize}
The model is optimised in three steps. First, we fit the normalisation of all sources within $25^{\circ}$ of {\ocen} (including {\ocen} itself), using a binned maximum likelihood analysis with the \textsc{optimise} routine \cite{wood} of %\textit{fermipy}
\textit{Fermitools} version~1.0.2.  
We quantify the significance of each detected point source via the test statistic
\begin{equation}
\mathrm{TS}\equiv 2(\mathrm{log} \mathcal{L}_{\mathrm{1}} - \mathrm{log} \mathcal{L}_{\mathrm{0}}), 
\end{equation}
where $\mathcal{L}_{1}$ and $\mathcal{L}_{0}$ are the maximum likelihood with and without the source in question (for one degree of freedom, the statistical significance is defined as $\sigma = \sqrt[]{\mathrm{TS}}$) \cite{like}.
Second, we remove all 31 (255) insignificant sources within $15^\circ$ ($25^\circ$) of {\ocen} with $\sigma<2$ %TS\,$<4$ 
or predicted number of photons $n_\gamma<2$. We then re-fit the normalisation of all 82 remaining point sources within $15^{\circ}$ of {\ocen}, and the spectral shape (i.e.\ spectral index and curvature) of those with $\sigma>5$. % TS\,$>25$. 
During this step we freeze the normalisation of sources $>15^{\circ}$ from {\ocen}. Third, we construct a $21^{\circ} \times 21^{\circ}$ binned TS map centered on {\ocen}, using the \textit{Fermitool} \textsc{gttsmap} routine. This works by calculating the increase in likelihood if a point source with spectral index 2 is placed sequentially in each $0.1^{\circ}$ square pixel. Within $15^{\circ}$ of {\ocen}, we find 6 new point sources of \textgamma-ray emission with $\sigma > 5$ that are not in the 4FGL. We add these point sources to our model, individually fitting their flux normalisation and power law spectral index (table~\ref{tab:NewSources}). We detect {\ocen} with $30\sigma$ (TS=$944$) significance, and separately refit its location.
A final TS map calculation shows no significant ($\sigma>5$) residual, indicating that we have successfully accounted for all nearby sources of \textgamma-ray emission (figure~\ref{fig:map}b).

The best-fit \textgamma-ray flux from {\ocen} is $1.13\pm0.05 \times 10^{-11}$\,ergs\,cm$^{-2}$\,s$^{-1}$. A likelihood analysis, using this best-fit flux model, finds {\ocen}'s best-fit location (RA $201.665^{\circ}$ Dec $-47.485^{\circ}$, with $1\sigma$ error radius $0.012^{\circ}$) 
is consistent with the 4FGL position of {\ocen}, and lies inside the core radius of its optical emission. 
A maximum likelihood analysis of its size indicates that, with \textit{Fermi}-LAT's point-spread function, its \textgamma-ray emission is statistically indistinguishable from a point source.
Splitting the data into 10 yearly bins (figure~\ref{fig:lc}) and comparing the likelihood profiles from each temporal bin to the overall likelihood profile from the 10-year mean flux, finds no evidence for either flux stochasticity or flux variability of the kind associated with standard astrophysical sources of \textgamma-rays.

%%%%%%%%%%%%%%%%%%%%%%%%%%%%%%%%%%%%%%%%%%%%%%%%%%%%%%%%%%%%%%%%%%%%%%%%%%%%%%%%%%%%%%
\section{Interpreting the gamma-ray emission as dark matter annihilation}

The DM-induced energy flux from a solid angle $\Delta\Omega$ is
\begin{equation}
E_{\gamma}^2 \dfrac{\mathrm{d}n}{\mathrm{d}E_{\gamma}}(\Delta \Omega, E_\gamma) = J(\Delta \Omega) \, \dfrac{\left\langle \sigma v \right\rangle}{\eta \,m_{\mathrm{DM}}^{2}} \, E_{\gamma}^2 \dfrac{\mathrm{d}N_{\gamma}}{\mathrm{d}E_{\gamma}}(E_{\gamma}, m_{\rm DM}),
\label{eqn:dndE}
\end{equation}
where $\rmd n/\rmd E_\gamma$ is the differential number of photons reaching the detector (per unit energy, sometimes known as $\rmd \Phi/\rmd E_\gamma$), the $J$-factor encodes the system's spatial morphology, $\left\langle \sigma v \right\rangle$ is the velocity-averaged annihilation cross section times relative velocity, $\eta = 2$ for self-annihilating DM, $m_{\mathrm{DM}}$ is the mass of the DM particle, and $\rmd N_\gamma/\rmd E_\gamma$ is the number of photons produced per annihilation.
To calculate $\rmd N_\gamma/\rmd E_\gamma$, we consider only one benchmark annihilation channel, into $b$ quarks (which, through hadronization and decay processes, produce photons)\footnote{This choice is motivated by the fact that the resulting \textgamma-ray spectrum closely resembles the shape of the observed spectral energy distribution. However, it would in principle be possible to introduce additional channels, with additional free parameters corresponding to the annihilation cross section into the various channels, and further improve the fit.}, based on the Monte Carlo particle physics event generator PYTHIA \cite{pythia}, and tabulated in Cirelli et al.\ (2011)\cite{Cirelli2011}.

Assuming that all \textgamma-ray emission from {\ocen} is due to DM annihilation, we fit the flux given in equation~\eqref{eqn:dndE} to the observed spectral energy distribution, with the DM particle mass $m_{\mathrm{DM}}$ and annihilation cross section $\left\langle \sigma v \right\rangle$ as free parameters. The two parameters act orthogonally: the particle mass essentially determines the energy of peak emission, and the cross section determines the flux normalisation. We calculate the statistical significance of these fits, including the full posterior probability distribution of the $J$-factor propagated from our analysis of stellar kinematics. The best-fit model has $\chi^2=6.8$ in $8$ degrees of freedom. 

%%%%%%%%%%%%%%%%%%%%%%%%%%%%%%%%%%%%%%%%%%%%%%%%%%%%%%%%%%%%%%%%%%%%%%%%%%%%%%%%%%%%%%
\section{Ruling out millisecond pulsars as the source of gamma-ray emission}

MSPs are unlikely to be responsible for the \textgamma-ray emission from {\ocen} for two reasons. First, the spectral energy distribution of the \textgamma-ray emission does not match that from unambiguously-identified MSPs in our Galactic neighbourhood. We fit the spectrum of nearby MSPs\cite{xingwang}, with a free flux normalisation. The best fit model fails to reproduce the falloff in flux at $E_\gamma<0.3$\,GeV, and achieves only $\chi^2=14$. Compared to the good fit for DM annihilation, this rules out the MSP spectral model with $p<0.01$. 

Second, MSPs also emit strongly in X-rays and radio waves -- but none have been confirmed at these wavelengths in {\ocen}. 
Observations with the \textit{Chandra X-ray Observatory}\cite{henleywillis2018} identified 40 `candidate' point sources in the region of X-ray colour-luminosity space ($0<2.5\log([0.5-1.5\mathrm{keV}]/[1.5-6\mathrm{keV}])<2$, and  $10^{30}<L_X(0.5-6\mathrm{keV})<10^{31}\,\mathrm{erg}\,\mathrm{s}^{-1}$) where MSPs reside \cite{heinke2005}. This region however, is not exclusively populated by MSPs. Already, optical observations confirmed one of these point sources to be a Cataclysmic Variable (CV) star, and four others to be ordinary Red Giant Branch stars or foreground stars unconnected with {\ocen} that do not emit in \textgamma-rays\cite{henleywillis2018}. Indeed, in the same region of colour space, globular cluster 47\,Tuc contains 118 point sources: 13\% MSPs, 4\% CVs, 28\% active binary stars, and 55\% others, which include main sequence stars, AGN, unresolved CVs/binary systems, and foreground stars\cite{heinke2005}.
If the same fraction of candidate sources in {\ocen}'s region of X-ray colour-luminosity space were MSPs, it would contain at most $5$. However, this is a gross upper limit, because the rate of stellar encounters, through which MSPs are created, is more than an order of magnitude lower in {\ocen} than in 47\,Tuc. The stellar encounter rate in {\ocen} is closest to that in globular cluster NGC6397, which contains just one MSP. 

If one or two MSPs did exist in {\ocen}, they would definitely have been observed at radio wavelengths. 
All 25 MSPs in 47\,Tuc (4.7\,kpc from Earth) have been confirmed in radio observations\cite{atnf}. 
In total, radio observations have confirmed 148 MSPs in globular clusters, including 101 more distant than {\ocen} and 92 at lower galactic latitudes (where interstellar scintillation broadens the pulsed emission outside the temporal windows used as a matched filter to gate the sensitivity of radio telescopes thus making it more difficult to observe the MSPs)\cite{pfreire}. 
In the Parkes Radio Telescope survey of globular clusters, of all clusters without MSPs, the deepest observations are of {\ocen}\cite{edwards2001}; still none were found\cite{possenti}.
Yet reproducing even the broad-band \textgamma-ray flux in {\ocen} would require $19\pm9$ average MSPs (assuming the average MSP spin-down power and an estimated average spin-down to \textgamma-ray luminosity conversion efficiency)\cite{abdo2010}. If such a great flux were concentrated in a few extremely bright MSPs, they would be even more impossible to miss.

%%%%%%%%%%%%%%%%%%%%%%%%%%%%%%%%%%%%%%%%%%%%%%%%%%%%%%%%%%%%%%%%%%%%%%%%

~
\newpage 
~\\
~

\section*{Supplementary material}
\setcounter{section}{0}
\setcounter{figure}{0}
\renewcommand{\thefigure}{S1}
%%%%%%%%%%%%%%%%%%%%%%%%%%

~

\begin{figure}[h]
\centering
\includegraphics[width=89mm]{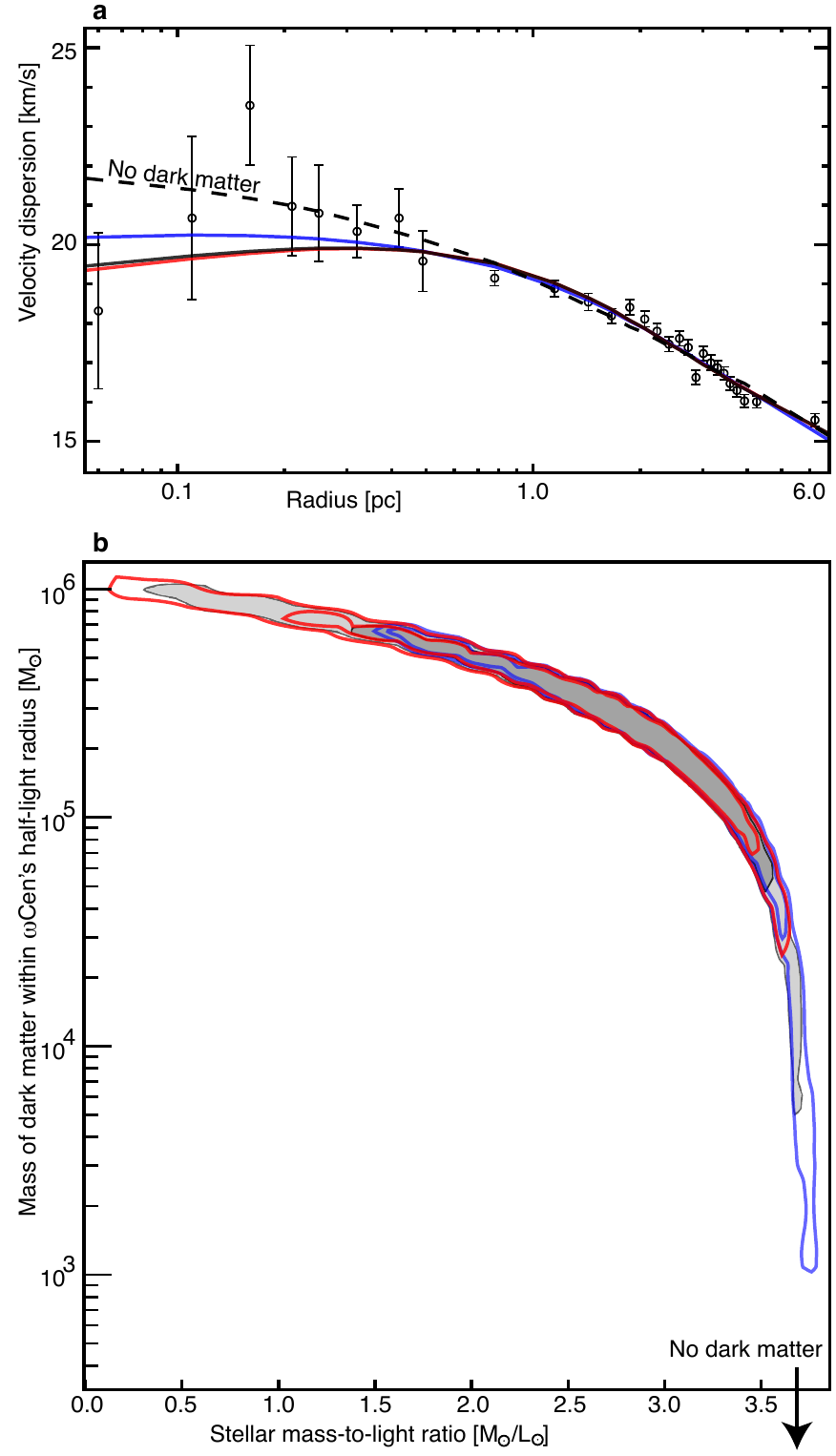}~
\includegraphics[width=89mm,trim={0 -26mm 0 0},clip]{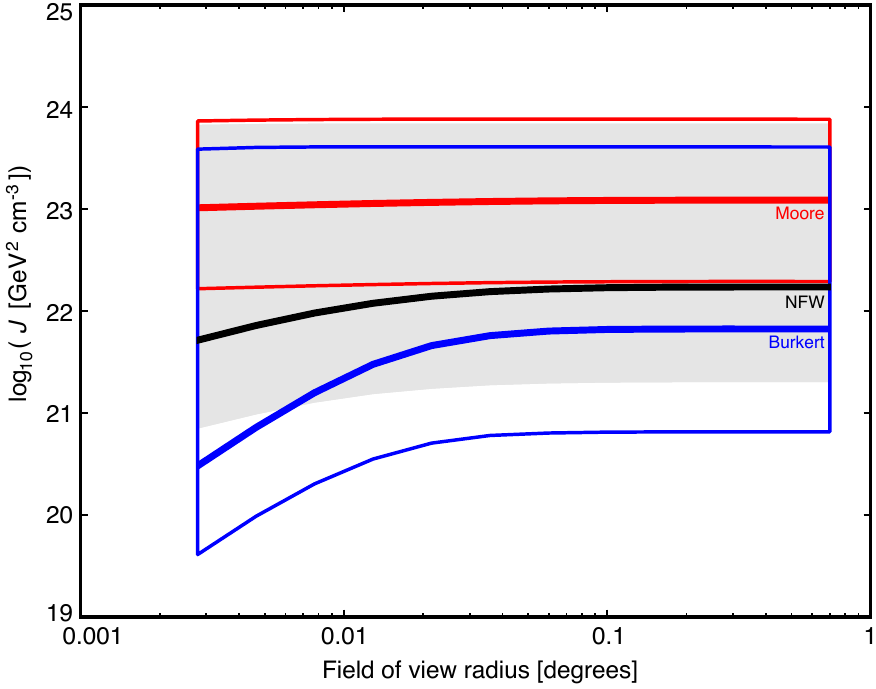}
\caption{As figures~\ref{fig:MLDM} and \ref{fig:jfactor}, but assuming no central black hole. In this scenario, we find a marginal increase in the statistical significance with which we rule out a model of {\ocen} that contains zero DM. However, the best-fit black hole mass was so low that excluding it does not significantly change the best-fit values of other parameters.}
\label{fig:MLDM_noBH}
\end{figure}


\newpage

\begin{thebibliography}{}
\bibitem{bert} Bertone, G. \& Tim, T.M.P., 2018, Nature, 562, 51
\bibitem{louie2018} Strigari L., 2018, Rept. Prog. Phys. 81, no.5, 056901
\bibitem{hooper2011} Hooper, D.\ \& Linden, T., 2011, PhRvD, 84, 123005
%\bibitem{goodenough2009} Goodenough, L.\ \& Hooper, D., 2009, ArXiv e-prints (arXiv:0910.2998) (2009). 0910.2998.
\bibitem{gordon2013} Gordon, C \& Macias, O, 2013, PhRvD, 88, 3521
\bibitem{leane2019} Leane, R.~K., \& Slatyer, T.~R.\ 2019, arXiv e-prints, arXiv:1904.08430
\bibitem{ackermann2017} Ackermann, M.\ et al., 2017, ApJ, 840, 43
\bibitem{deboer2017} de Boer, W., Bosse, L., Gebauer, I., Neumann, A. \& Biermann, P., 2017, Phys. Rev. D 96, 043012
%\bibitem{han2012} Han J., Frenk C., Eke V. et al. 2012, MNRAS 427, 1651
%\bibitem{yuan2014} Yuan, Q. \& Zhang, B., 2014, J. High Energy Astrophys. 3, 1–8
\bibitem{macias2018} Macias, O., Gordon, C., Crocker, R., Coleman, B., Paterson, D., Horiuchi, S.\ \& Pohl, M., 2018, Nature Astronomy, 2, 387
%\bibitem{acker2015} Ackermann, M. et al., \textit{Fermi}-LAT Collaboration, 2015, Phys.\ Rev.\ Lett.\ 115, 231301
%\bibitem{DW2015} Drlica-Wagner, A., et al. (Fermi-LAT Collaboration, DES Collaboration), 2015, Astrophys. J. 809, L4
\bibitem{albert2017} Albert, A.\ et al., 2017, ApJ, 834, 110
%\bibitem{myeong2018} Myeong, G.C., et al., 2018, MNRAS, 478, 5449
%\bibitem{vandeven2006} van de Ven, G., van den Bosch, R.G.E., Verolme, E.K. \& de Zeeuw, P.T., 2006, A\&A, 445, 513
%\bibitem{norris1996} Norris, J.E., Freeman, K.C. \& Mighell, K.J., 1996, ApJ, 462, 241
\bibitem{lee1999} Lee, Y.-W., Joo J.-M., Sohn Y.-J., Rey S.-C., Lee H.-C.\& Walker, A.R., 1999, Nature, 402, 55
%\bibitem{hilker2000} Hilker, M. \& Richtler, T., 2000, A\&A, 362, 895
\bibitem{piotto2005} Piotto, G., et al., 2005, ApJ, 621, 777
\bibitem{sollima2005} Sollima, A., Pancino, E., Ferraro, F.R., Bellazzini, M., Straniero, O. \& Pasquini, L., 2005, ApJ, 634, 332
\bibitem{ibata2019} Ibata, R.A., Bellazzini, M., Malhan, K., Martin, N. \& Bianchini, 2019, Nature Astronomy, tmp 258l
\bibitem{bekki2003} Bekki K. \& Freeman K. C., 2003, MNRAS, 346, L11
%\bibitem{bekki2006} Bekki, K., \& Norris, J.E., 2006, ApJ, 637, L109
%\bibitem{boyles2011} Boyles, J., Lorimer, D. R., Turk, P. J., Mnatsakanov, R., Lynch, R. S., Ransom, S. M., Freire, P. C. \& Belczynski, K., 2011, ApJ, 742, 51B
\bibitem{henleywillis2018} Henleywillis, S., Cool, A.M., Haggard, D., Heinkie, C., Callanan, P. \& Zhao, 2018, MNRAS, 479, 2834
\bibitem{edwards2001} Edwards, R. T., van Straten, W. \& Bailes, M., 2001, ApJ, 560, 365
\bibitem{possenti} Possenti, A., D'Amico, N., Corongiu, A., Manchester, D., Sarkissian, J., Camilo, F.\& Lyne, A., 2005, ASPC, 328, 189
\bibitem{steigman2012} Steigman, G., Dasgupta, B., \& Beacom, J.~F.\ 2012, PRD, 86, 023506
\bibitem{baumgardt2017} Baumgardt, H., 2017, MNRAS, 464, 2174
\bibitem{harris1996} Harris, W.E. 1996, AJ, 112, 1487
\bibitem{dinescu1999} Dinescu, D.I., Girard, T.M., van Altena, W.F., 1999, AJ, 117, 1792
\bibitem{Weldrake2007} Weldrake, D., Sackett, P. \& Bridges, T., 2007, Astronomical Journal, 133, 1447
\bibitem{vdM2010} van der Marel, R.~P., \& Anderson, J.\ 2010, \apj, 710, 1063
\bibitem{Tolstoy2009} Tolstoy, E., Hill, V. \& Tosi, M., 2009, ARA\&A, 2009, 47, 371
\bibitem{dsouza2018} D'Souza, R. \& Bell, E., 2018, Nature Astronomy, 2, 737
\bibitem{Abadi2010} Abadi, M., Navarro, J. \& Fardal, M., 2010, MNRAS, 407, 435
\bibitem{bahramian2013} Bahramian, A., Heinke, C.O., Sivaokoff, G.R. \& Gladstone, J.C., 2013, ApJ, 766, 136
\bibitem{santana2013} Santana, F.A., Munoz, R.R., Geha, M, Cote, P., Stetson, P., Simon, J.D. \& Djorgovski, S.G., 2013, ApJ,774, 106 
%\bibitem{gendre2003} Gendre, B., Barret, D. \& Webb, N., 2003, A\&A 400, 521
\bibitem{cv} Ackermann M. et al., 2014, Science, 345, 554
%\bibitem{morris2017} Morris, P.J., Cotter, G., Brown, A.M. \& Chadwick, P.M., 2017, MNRAS, 465, 1218
\bibitem{nerenov2008} Neronov, A., \& Aharonian F.A., 2008, ApJ, 671, 85
\bibitem{Boddy2018} Boddy, K.~K., Kumar, J., \& Strigari, L.~E., 2018, PRD, 98, 063012
\bibitem{Kamann2018} Kamann, S., Husser, T.-O., Dreizler, S, et al.\ 2018, MNRAS, 473, 5591 
\bibitem{Baumgardt2018} Baumgardt, H.\ \& Hilker, M., 2018, MNRAS, 478, 1520
\bibitem{Baumgardt2019} Baumgardt, H., Hilker, Solima, A. \& Bellini, A., 2019, MNRAS, 482, 5138 
\bibitem{Watkins2015} Watkins, L.~L., van der Marel, R.~P., Bellini, A., et al.\ 2015, \apj, 803, 29
\bibitem{Walker2011} Walker, M. \& Pe{\~n}arrubia, J., 2011, ApJ, 742, 20
\bibitem{barber2016} Barber, C., Schaye, J., Bower, R., Crain, R., Schaller, M. \& Theuns, T., 2016, MNRAS, 460, 1147
\bibitem{Chua2019} Chua, K.~T.~E., Pillepich, A., Vogelsberger, M. \& Hernquist, L., 2019, MNRAS, 484, 476
\bibitem{Penarrubia2009} Penarrubia, J., Navarro, J. \& McConnachie, A., 2009, ApJ, 698, 222
\bibitem{Helmi2018} \textit{Gaia} Collaboration, Helmi, A., et al.\ 2018, A\&A, 616, A12
\bibitem{nfw1996} Navarro, J., Frenk, C.\ \& White, S., 1996, ApJ, 462, 563
\bibitem{moore1999} Moore, B., Ghigna, S., Governato, F., Lake, G., Quinn, T., Stadel, J.\ \& Tozzi, P., 1999, ApJL, 524, 19
\bibitem{burkert1995} Burkert, A., 1995, ApJL, 447, 25
\bibitem{multinest} Feroz, F., Hobson, M.~P., \& Bridges, M.\ 2009, MNRAS, 398, 1601
\bibitem{Conroy2009} Conroy, C., Gunn, J.\ \& White, M., 2009, ApJ, 699, 486
\bibitem{Wolf2010} Wolf, J., Martinez, G., Bullock, J., et al.\ 2010, MNRAS, 406, 1220
\bibitem{McConnachie2012} McConnachie, A., 2012, AJ, 144,4
\bibitem{Fattahi2018} Fattahi, A., Navarro, J., Frenk, C., et al. \& 2018, MNRAS, 476, 3816
\bibitem{Mieske2008} Mieske, S., Hilker, M., Jord{\'a}n, A., et al.\ 2008, A\&A, 487, 921
\bibitem{Woo2008} Woo, J., Courteau, S. \& Dekel, A., 2008, MNRAS, 390, 1453  
\bibitem{Sofue2013} Sofue, Y., 2013, PASJ, 65, 118
\bibitem{Ludlow2016} Ludlow, A., Bose, S., et al.\ 2016, 460, 1214
\bibitem{Howley2013} Howley, K.~M., Guhathakurta, P., van der Marel, R., et al.\ 2013, ApJ, 765, 65
\bibitem{Goerdt2008} Goerdt, T., Moore, B., Kazantzidis, S., et al.\ 2008, MNRAS, 385, 2136
%\bibitem{katsoulakos2018} Katsoulakos, G.\ \& Rieger, F., 2018, ApJ, 852, 112
\bibitem{brown2018} Brown, A.M., Lacroix, T., Lloyd, S., Boehm, C. \& Chadwick, P., 2018, PhRvD, 98, 1301 
\bibitem{4fglcat} \textit{Fermi}-LAT Collaboration, 2019, arxiv:1902.10045
\bibitem{wood} Wood, M., Caputo, R., Charles, E., et al.\ 2017, International Cosmic Ray Conference, 35, 824
\bibitem{like} Mattox, J.R., et al., 1996, ApJ, 461, 396
\bibitem{pythia} Sj{\"o}strand T., Mrenna S., Skands P., 2008, CoPhC, 178, 852
\bibitem{Cirelli2011} Cirelli, M., Corcella, G., Hektor, A., et al.\ 2011, JCAP, 2011, 051
\bibitem{xingwang} Xing, Y. \& Wang, Z., 2016, \apj, 831, 143
\bibitem{heinke2005} Heinke, C.O., Grindlay, J.E., Edmonds, P.D., Cohn, H.N., Lugger, P.M., Camilo, G., Bogdanov, S. \& Freire, P.C., 2005, ApJ, 625, 796
\bibitem{atnf} \url{http://www.atnf.csiro.au/research/pulsar/psrcat/}
\bibitem{pfreire} Freire, P., 2018, \url{http://www3.mpifr-bonn.mpg.de/staff/pfreire/GCpsr.html}
\bibitem{abdo2010} Abdo, A.A., et al., 2010, A\&A, 524, 75
%\bibitem{Blumenthal1986} Blumenthal, G., Faber, S. \& Flores, R., 1986, ApJ, 301, 27
%\bibitem{dist} Braga, V.F., Stetson, P.~B., Bono, G., et al.\ 2018, AJ, 155, 137 



\end{thebibliography}
\end{document}